\documentclass[12pt,reqno]{amsart}
\usepackage{epsfig}
\usepackage{graphicx}
\begin{document}
\baselineskip=0.8cm 
\theoremstyle{plain}
\newtheorem{thm}{Theorem}[section]
\newtheorem{lem}{Lemma}[section]
\newtheorem{prop}{Proposition}[section]
\newtheorem{coll}{Conclusion}
\theoremstyle{remark}
\newtheorem{rem}{Remark}[section]
\title{Darboux Transformation and Variable Separation Approach:
 the Nizhnik-Novikov-Veselov equation}
\author{Heng-chun Hu$^1$, Sen-yue Lou$^{1,2}$ and Qing-ping Liu$^3$}
\dedicatory{{$^1$Department of Physics, Shanghai Jiao Tong
University, Shanghai, 200030, P.R.China}\\{$^2$Department of
Physics, Ningbo University, Ningbo, 315211, P.R.China}\\
$^3$Beijing Graduate School, China University of Mining and
Technology, Beijing 100083, P. R. China}
\begin{abstract}
Darboux transformation is developed to systematically find
variable separation solutions for the Nizhnik-Novikov-Veselov
equation. Starting from a seed solution with some arbitrary
functions, the once Darboux transformation yields the variable
separable solutions which can be obtained from the truncated
Painlev\'e analysis and the twice Darboux transformation leads to
some new variable separable solutions which are the generalization
of the known results obtained by using a guess ansatz to solve the
generalized trilinear equation.

\vskip.2in

\leftline{\bf \emph{PACS.}02.30.Jr, 02.30.Ik, 05.45.Yv.}
\end{abstract}

\maketitle
To find some exact solutions of integrable systems has focused
many mathematicians and physicists' attention since the soliton
theory came into being at the 1960's. There are many important
methods to obtain the special solutions of a given  soliton
equation. Some of the most important methods are the inverse
scattering transformation (IST) approach\cite{GGKM}, the bilinear
form\cite{Hirota}, symmetry reduction\cite{Olver}, B\"acklund
transformation and Darboux transformation \cite{Liu} etc. In
comparison with the linear case, it is known that IST is an
extension of the Fourier transformation in the nonlinear case. In
addition to the Fourier transformation, there is another powerful
tool called the variable separation method \cite{Lou} in the
linear case. Recently, some kinds of variable separation
approaches had been developed to find new exact solutions of
nonlinear models, say, the classical method, the differential
St\"ackel matrix approach  \cite{Miller}, the geometrical method
\cite{Do1}, the ansatz-based method \cite{Rz1, Do1}, functional
variable separation approach\cite{Zh1}, the derivative dependent
functional variable separation approach\cite{Zsl}, the formal
variable separation approach (nonlinearization of the Lax pairs or
symmetry constraints) \cite{Lou3} and the informal variable
separation methods \cite{Lou1}--\cite{Lou4}.

Especially, for various (2+1)-dimensional nonlinear physics
models, a quite universal formula \begin{equation}\label{0}
U\equiv \frac{2(a_1a_2-a_0a_3)q_{y}p_{x}}{(a_0+a_1 p+a_2 q+a_3 p
q)^2},
\end{equation}
where $a_i, \ (i=0,\ 1, 2,\ 3)$ are arbitrary constants and
$p=p(x,\ t)$ and $q=q(y,\ t)$ are arbitrary constants of the
indicated variables, is found by using the informal variable
separation approach\cite{Lou1}--\cite{Lou4}. Starting from the
universal formula \eqref{0}, abundant localized excitations like
the dromions, lumps, ring solitons, breathers, instantons,
solitoffs, fractal and chaotic patterns are found. Now a very
important question is can we find the universal formula from other
well known powerful methods like the IST approach, dressing
method, Darboux transformation (DT) etc?

DT is one of the most powerful methods to construct a broad class
of considerable physical interest and important nonlinear
evolution equations such as the well-known Korteweg-de Vries (KdV)
equation, the Kadomtsev-Petviashvili (KP) equation, the
Davey-Stewartson (DS) equation, the  sine-Gordon (SG) equation
\cite{Matveev} and so on. In this letter, we use the Darboux
transformation to study the variable separable solutions for the
(2+1) dimensional Nizhnik-Novikov-Veselov system.\cite{NNV}
\begin{equation}\label{NNVu}
u_t=u_{xxx}+u_{yyy}+3(vu)_x+3(uw)_y,
\end{equation}
\begin{equation}\label{NNVvw}
u_x=v_y,\qquad u_y=w_x.
\end{equation}
The (2+1)-dimensional NNV equation is an only known isotropic Lax
integrable extension of the well-known (1+1)-dimensional KdV
equation. Many authors have studied the solutions of the NNV
equation. For example, Boiti et al\cite{BLM} solved the NNV
equation via the IST; Tagami and Hu and Li obtained the
soliton-like solutions of the NNV equation by means of B\"acklund
transformation\cite{Hu1}; Hu also gave out the nonlinear
superposition formula of the NNV equation\cite{Hu2}; Some special
types of multi-dromion solutions were found by Radha and
Lakshmanan\cite{Radha}; The generalized localized excitations
expressed by \eqref{0} were given in \cite{Lou1} and \cite{Lou4}
and the special binary Darboux transformation was given in
\cite{Matveev}.

It is known that the NNV equation system \eqref{NNVu} and
\eqref{NNVvw} can be represented as a compatibility condition of
the linear system
 \begin{equation}\label{LaxX}
 \Phi_{xy}+u\Phi=0,
 \end{equation}
 \begin{equation}\label{LaxT}
 \Phi_t=\Phi_{xxx}+\Phi_{yyy}+3v\Phi_x+3w\Phi_y.
 \end{equation}
In general, in order to construct the solutions of a given
equation by means of Darboux transformation, one may use a fixed
solution for a special spectrum parameter of the Lax pair. Then
one can get a new solution from an old one with Darboux
transformation. Without spectral parameter in the Lax pair of the
NNV equation, we have to construct a binary Darboux transformation
for the NNV equation. At the same time, with the variable
separated approach, two arbitrary functions can be entered into
the solution.

It is straightforward to see that the NNV system \eqref{NNVu} and
\eqref{NNVvw} possesses the following trivial solution
\begin{equation}\label{uvw0}
u=0,\ v=v_0(x,\ t),\ w=w_0(y,\ t),
\end{equation}
where $v_0(x,\ t)$ and $w_0(y,\ t)$ are arbitrary functions of
$\{x,\ t\}$ and $\{y,\ t\}$ respectively.

In order to find some new solutions via Darboux transformation and
the seed solution \eqref{uvw0}, the key step is to find the fixed
solution of the Lax pair \eqref{LaxX} and \eqref{LaxT} with the
seed \eqref{uvw0}.

It is evident that Eq.\eqref{LaxX} have a variable separation
solution in the form
\begin{equation}\label{phi0}
\Phi_0=p+q
\end{equation}
where $p=p(x,t)$, $q=q(y,t)$ are two arbitrary functions when
$u=0$. Substituting Eq. \eqref{phi0} and Eq. \eqref{uvw0} into Eq.
\eqref{LaxT} yields
\begin{equation}\label{pqt}
p_t+q_t=p_{xxx}+q_{yyy}+3v_0p_x+3w_0q_y.
 \end{equation}
It is clear that Eq. \eqref{pqt} can be solve by the usual
variable separable approach and the result reads
\begin{equation}\label{pt}
p_t=p_{xxx}+3v_0p_x+c(t),
 \end{equation}
\begin{equation}\label{qt}
q_t=q_{yyy}+3w_0q_y-c(t)
 \end{equation}
where $c(t)$ is an arbitrary function of $t$. For given functions
$v_0,\ w_0$ and $c(t)$, it is still difficult to solve \eqref{pt}
and \eqref{qt}. However, because of the arbitrariness of the
functions $v_0$ and $w_0$, we can solve the problem in an
alternative way. If we consider the functions $p$ and $q$ as
arbitrary functions, then $v_0$ and $w_0$ can be solved out from
\eqref{pt} and \eqref{qt}
\begin{equation}\label{v0}
v_0=(3p_{x})^{-1}(p_{t}-p_{xxx}-c(t))
\end{equation}
\begin{equation}\label{w0}
w_0=(3q_{y})^{-1}(q_{t}-q_{yyy}+c(t)).
\end{equation}

As usual, the binary Darboux  transformation for the NNV equation
can be constructed by introducing the closed 1-form
 \begin{eqnarray}
 \omega(\Phi,\Phi_0)&=&(\Phi\Phi_{0x}-\Phi_x\Phi_0)dx-(\Phi\Phi_{0y}-\Phi_0\Phi_y)dy+
 \{\Phi(\Phi_{0xxx}-\Phi_{0yyy})\nonumber\\
&&+\Phi_0(\Phi_{yyy}-\Phi_{xxx})+2(\Phi_{xx}\Phi_{0x}
-\Phi_x\Phi_{0xx}+\Phi_{0yy}\Phi_{y}
-\Phi_{0y}\Phi_{yy})\nonumber\\
&&+3v(\Phi\Phi_{0x}-\Phi_x\Phi_0)+3w(\Phi_0\Phi_{y}-\Phi_{0y}\Phi)\}dt.\label{1form}
\end{eqnarray}
Then the new wave function is
\begin{eqnarray}\label{phi1}
\Phi[1]=\Phi_0^{-1}\int\omega
\end{eqnarray}
and the equation system Eq.\eqref{LaxX} and \eqref{LaxT} is
covariant with respect to the transformation \eqref{phi1} and the
transformed coefficients $u[1]$, $v[1]$ and $w[1]$ are
\begin{eqnarray}\label{u1}
u[1]&=&2(\ln\Phi_0)_{xy}=\dfrac{-2p_xq_y}{(p+q)^2},\\
v[1]&=&v_0+2(\ln\Phi_0)_{xx}=v_0+\dfrac{2p_{xx}}{p+q}-\dfrac{p_x^2}{(p+q)^2},\label{v1}\\
w[1]&=&w_0+2(\ln\Phi_0)_{yy}=w_0+\dfrac{2q_{yy}}{p+q}-\dfrac{q_y^2}{(p+q)^2},\label{w1}
\end{eqnarray}
where $v_0$ and $w_0$ determined by Eqs.\eqref{v0} and \eqref{w0}.
From \eqref{u1}, we see that the solution $u[1]$ obtained by the
once Darboux transformation is only a special case of the
universal expression \eqref{0} with $a_3=0$. The solution
\eqref{u1}--\eqref{w1} can also be obtained by the truncated
Painlev\'e expansion using the similar method as for the AKNS
system\cite{LLT}.

By means of the iteration of the Darboux transformation, we can
construct the second Darboux transformation from the first Darboux
transformation. We also take two special solutions of the Lax pair
\eqref{LaxX} and \eqref{LaxT} as the variable separable ones,
$\Phi_1=p_1(x,t)+q_1(y,t)$ and $\Phi_2=p_2(x,t)+q_2(y,t)$, then
the transformed wave function is
\begin{equation}\label{phi2}
\Phi[2]=\int \omega(\Phi_1,\Phi_2)
\end{equation}
which leads to the formula
\begin{eqnarray}
u[2]&=&2(\ln\int\omega(\Phi_1,\Phi_2))_{xy},\label{u2}\\
v[2]&=&v_0+2(\ln\int\omega(\Phi_1,\Phi_2))_{xx},\label{v2}\\
w[2]&=&w_0+2(\ln\int\omega(\Phi_1,\Phi_2))_{yy}\label{w2}
\end{eqnarray}
According to the selection for the arbitrary functions, the
potential functions $v_0$ and $w_0$ are related to the functions
$p_1,\ p_2,\ q_1$ and $q_2$ by
\begin{eqnarray}
p_{1t}-p_{1xxx}-c_1(t)-v_0p_{1x}=0,\label{v0p1}\\
q_{1t}-q_{1yyy}+c_1(t)-w_0q_{1y}=0,\label{w0q1}\\
p_{2t}-p_{2xxx}-c_2(t)-v_0p_{2x}=0,\label{v0p2}\\
q_{2t}-q_{2yyy}+c_2(t)-w_0q_{2y}=0. \label{w0q2}
\end{eqnarray}
Now we can consider one of the functions $p_1=p_1(x,t)$ and
$p_2=p_2(x,t)$ as an arbitrary function of $\{x,\ t\}$ and one of
the functions $q_1=q_1(y,t)$ and $q_2=q_2(y,t)$ as an arbitrary
function of $\{y,\ t\}$ while the remained functions of $v_0,\
w_0,\ p_1,\ p_2,\ q_1$ and $q_2$ should be determined by
\eqref{v0p1}, \eqref{w0q1}, \eqref{v0p2} and \eqref{w0q2}.

It is not easy to solve the equation system \eqref{v0p1},
\eqref{w0q1}, \eqref{v0p2} and \eqref{w0q2}. However, if we choose
\begin{eqnarray}
q_1=0,\qquad p_2=0,\label{q1p2},
\end{eqnarray}
then the Eqs. \eqref{u2}-\eqref{w2} can be simplified into
\begin{eqnarray}
u[2]&=&2\ln(c+p_1 q_2)_{xy},\label{ru2}\\
v[2]&=&v_0+2\ln(c+p_1 q_2)_{xx},\label{rv2}\\
w[2]&=&w_0+2\ln(c+q_1 q_2)_{yy},\label{rw2}
\end{eqnarray}
where $c$ is an arbitrary integral constant, $p_1$ and $q_2$ are
arbitrary functions of $\{x,\ t\}$ and $\{y,\ t\}$ while $v_0$ and
$w_0$ are fixed by \eqref{v0p1} and \eqref{w0q2}. It is easy to
prove that the results \eqref{ru2}--\eqref{rw2} are equivalent to
those obtained via the usual variable separation of the
multi-linear equation\cite{Lou1}. Actually, by re-writing $c,\
p_1$ and $q_2$ as
\begin{eqnarray}\label{cpq}
c=-a_0-\dfrac{a_1a_2}{a_3},\ p_1=a_2+a_3p,\
q_2=q+\dfrac{a_1}{a_3},
\end{eqnarray}
the negative value of the right hand side of \eqref{ru2} is
transformed to the universal quantity expressed by \eqref{0}.

Various types of coherent localized structures for the physical
field $u$ of the NNV system had been described in \cite{Lou1} and
\cite{Lou4} thanks to the arbitrariness of the functions $p$ and
$q$.
We do not repeated these known localized excitations but
write down only one new type of localized solutions for the field
$u$ expressed by \eqref{ru2} with \eqref{cpq} to complement the
results of \cite{Lou4}.

In addition to the continuous localized excitations in
(1+1)-dimensional nonlinear systems, some types of significant
weak solutions like the peakons\cite{CH, peakon} and compactons
\cite{compacton} have been attract much attention of both
mathematicians and physicists. In \cite{Lou4} and \cite{peakon},
the possible (2+1)-dimensional localized peakons had been given.
The (1+1)-dimensional compacton solutions which describes the
typical (1+1)-dimensional soliton solutions with finite wavelength
when the nonlinear dispersion effects were firstly given by
Rosenau et al.\cite{Rose1} and may have many interesting
properties and possible physical
applications\cite{compacton}--\cite{LouWu}. For instance, the
compacton equations may be used to study the motion of
ion-acoustic waves and a flow of a two layer liquid\cite{Rose2}.
In \cite{LouWu}, the Painlev\'e integrability of two sets of
Korteweg-de Vries (KdV) type and modified KdV type compacton
equations are proved. Because of the entrance of arbitrary
functions in the (2+1)-dimensional nonlinear physics models, it is
easy to find some types of multiple compacton solutions by
selecting the arbitrary functions appropriately. For instance, if
we fixed  the functions $p$ and $q$ as
\begin{eqnarray}\label{compp}
p=\sum_{i=1}^N \left\{\begin{array}{ll} 0 & x+c_it\leq x_{0i}-\frac{\pi}{2k_i}\\
    b_i\sin(k_i(x+c_it-x_{0i}))+b_i & x_{0i}-\frac{\pi}{2k_i}<x+c_it\leq x_{0i}+\frac{\pi}{2k_i}\\
    2b_i & x+c_it>x_{0i}+\frac{\pi}{2k_i}
\end{array}\right. ,
\end{eqnarray}
and
\begin{eqnarray}\label{compq}
q=\sum_{j=1}^M \left\{\begin{array}{ll} 0 & y\leq y_{0j}-\frac{\pi}{2l_j}\\
    d_j\sin(l_i(y-y_{0j})) +d_j &  y_{0j}-\frac{\pi}{2l_j}<y\leq y_{0j}+\frac{\pi}{2l_j}\\
    2d_j & y>y_{0j}+\frac{\pi}{2l_j}
\end{array}\right. .
\end{eqnarray}
where $b_i, \ k_i,\ x_{0i},\ d_j,\ l_j$ and $y_{0j}$ are all
arbitrary constants, then the solution \eqref{ru2} with
\eqref{cpq} becomes a multi-compacton solution.

From \eqref{compp} and \eqref{compq}, one can see that the
piecewise functions $p$ and $q$ of the compacton solutions are
once differentiable
\begin{eqnarray}\label{comppx}
p_x=\sum_{i=1}^N \left\{\begin{array}{ll} 0 & x+c_it\leq x_{0i}-\frac{\pi}{2k_i}\\
    b_ik_i\cos(k_i(x+c_it-x_{0i})) & x_{0i}-\frac{\pi}{2k_i}<x+c_it\leq x_{0i}+\frac{\pi}{2k_i}\\
    0 & x+c_it>x_{0i}+\frac{\pi}{2k_i}
\end{array}\right. ,
\end{eqnarray}
and
\begin{eqnarray}\label{compqy}
q_y=\sum_{j=1}^M \left\{\begin{array}{ll} 0 & y\leq y_{0j}-\frac{\pi}{2l_j}\\
    d_jl_j\cos(l_j(y-y_{0j})) &  y_{0j}-\frac{\pi}{2l_j}<y\leq y_{0j}+\frac{\pi}{2l_j}\\
    0 & y>y_{0j}+\frac{\pi}{2l_j}
\end{array}\right. ,
\end{eqnarray}
So the multi-compacton solutions expressed by \eqref{0} (i.e.,
\eqref{ru2} with \eqref{cpq}) are still continuous any where
though there are some isolated non-differentiable lines.

In (1+1)-dimensions, a non-differentiable solution, $u=u_0$, like
the the compacton (and/or peakon) is called a weak solution of a
nonlinear (1+1)-dimensional PDE (partial differential equation)
\begin{eqnarray}\label{pde}
F(u,u_x,u_t,u_{xx},...)\equiv F(u)=0
\end{eqnarray}
under the meaning that though the compacton (and/or peakon)
solution is non- differentiable at some points $x=x_i(t)$, the
distribution, $F(u_0)=f(\delta(x-x_i(t))$ (where
$\delta(x-x_i(t))$ is a Dirac $\delta$ function and $f$ is a
function of the Dirac $\delta$ function and its derivatives) is
really a zero distribution that means $\int_{-\infty}^{+\infty}
f\psi(x,t)=0$ for arbitrary $\psi$.

However, in (2+1)-dimensions, the (2+1)-dimensional compactons
(and the peakons reported in \cite{Lou4}) are exact solutions of
some (2+1)-dimensional nonlinear equations are guaranteed by the
variable separation procedure. When we substituting the piecewise
solutions into the nonlinear PDEs, the Dirac delta function(s) in
$x$ ($y$) directions will be vanished by the differential operator
in other direction, $\partial_y$ ($\partial_x$). In other words,
we can take (have taken) $\partial_y \psi(x,t) =0$ no matter the
function $\psi$ is a continuous function of $\{x,t\}$ or a
generalized distribution functions (with some Dirac delta
functions) of $\{x,t\}$.

Fig.1 is the evolution plot of a three compacton solution
\eqref{ru2} with \eqref{cpq}, \eqref{compp}, \eqref{compq} and
\begin{eqnarray}\label{2ct}
&&N=3,\ M=1,\ a_0=20,\ a_1=a_2=25a_3=1,\ b_1=b_3=-c_2=c_3=-2, \nonumber \\
&&-b_2=-c_1=d_1=k_{1}=k_{2}=k_3=l_{1}=1, \
x_{01}=x_{02}=x_{03}=y_{01}=0.
\end{eqnarray}

\input epsf
\begin {figure}
\centering \epsfxsize=7cm\epsfysize=5cm\epsfbox{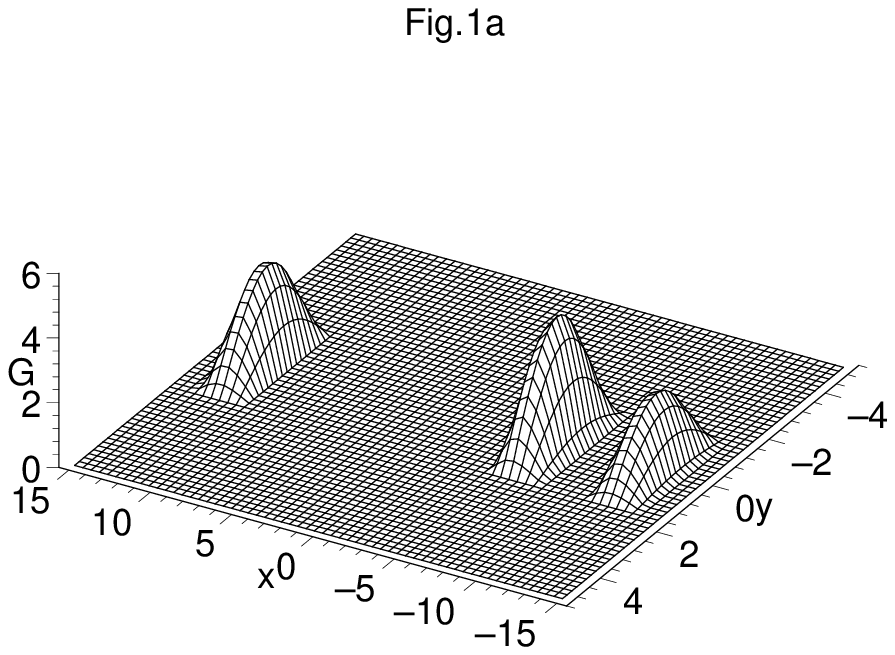}
\epsfxsize=7cm\epsfysize=5cm\epsfbox{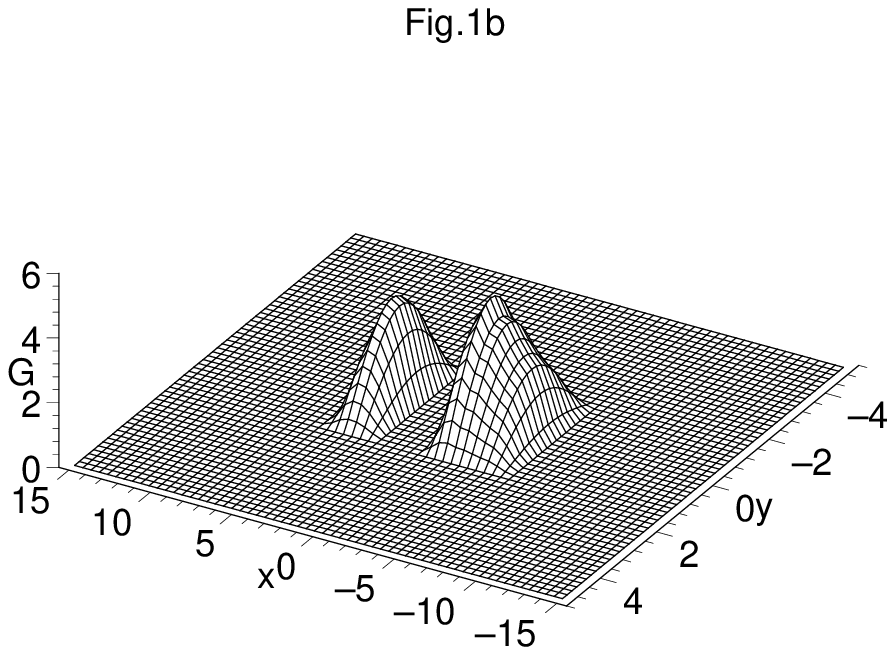}
\epsfxsize=7cm\epsfysize=5cm\epsfbox{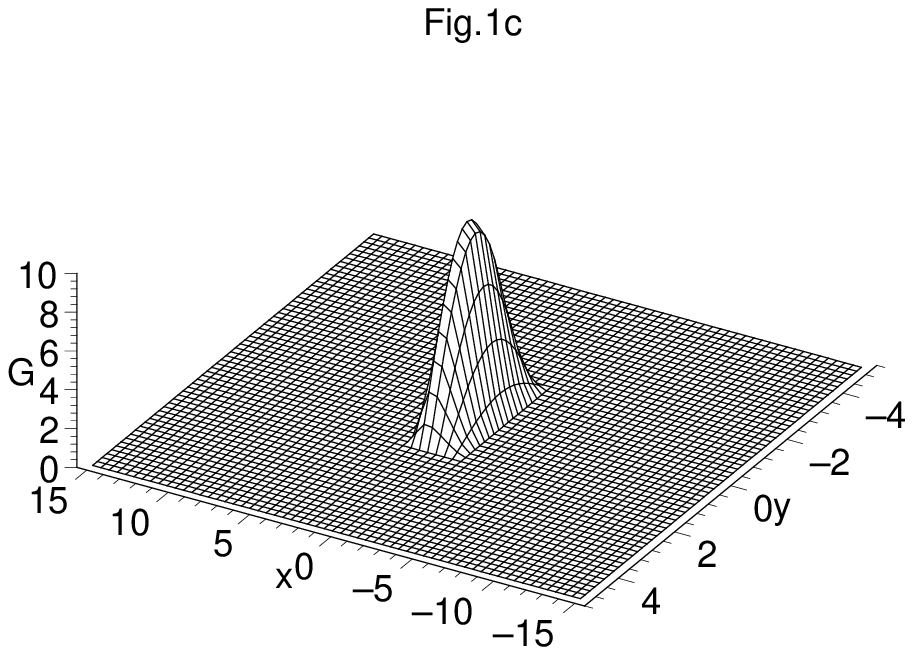}
\epsfxsize=7cm\epsfysize=5cm\epsfbox{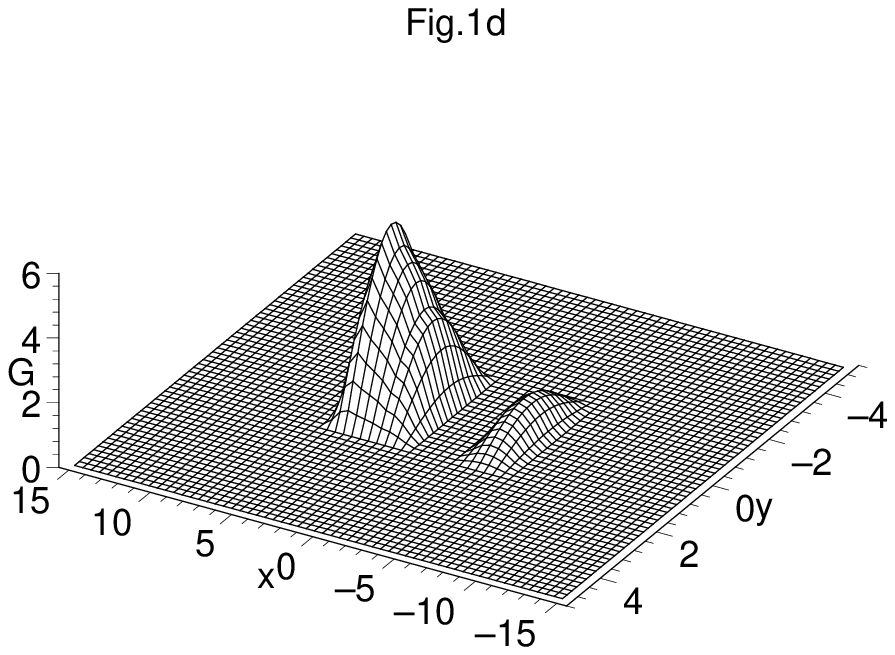}
\epsfxsize=7cm\epsfysize=5cm\epsfbox{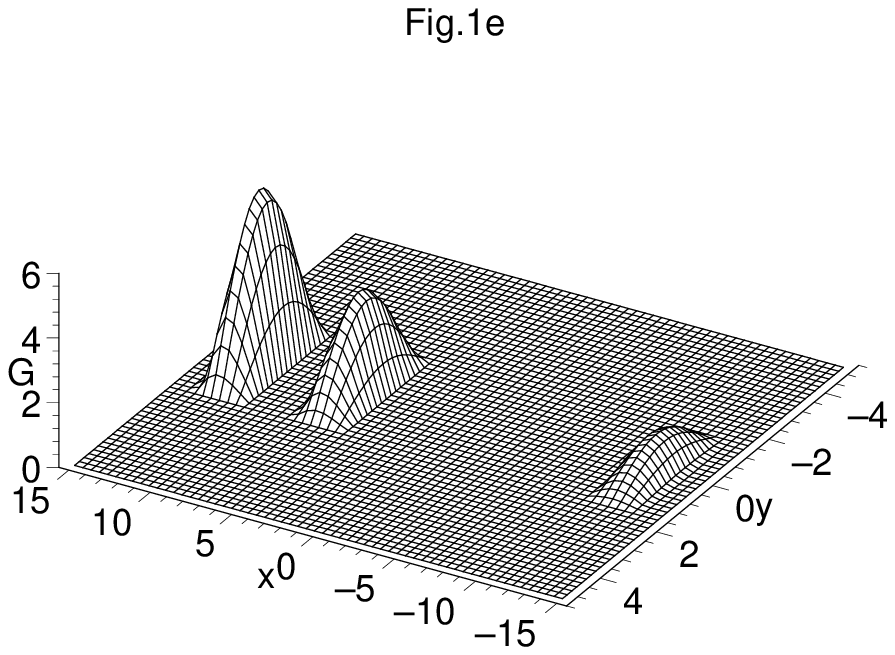} \caption{Evolution
plot for the quantity $G\equiv -1000 u[2]$ related to the three
compacton solution \eqref{ru2} with \eqref{cpq}, \eqref{compp},
\eqref{compq} and \eqref{2ct} at times (a) $t=-6$, (b) $t=-2$, (c)
$t=0$, (d) $t=3$, (e) $t=6$.}
\end{figure}

In \cite{Lou4}, we have pointed out that (i) the interaction
between two travelling ring shape soliton solutions is completely
elastic and (ii) the interaction between two travelling peakons is
not completely elastic, two peakons may completely exchange their
shapes. From Fig. 1, we see that the interaction between two
compactons exhibits a new phenomenon. The interaction is
nonelastic but two compactons do not completely exchange their
shapes.

Because of the arbitrariness of the functions $p$ and $q$ may have
also quite rich structures. For instance, (2+1)-dimensional
compactons may be only compacted at one direction. For conveniece
later we call this type of compactons the partial compactons. Fig.
2 is a plot of a special partial compacton solution \eqref{ru2}
with \eqref{cpq}, \eqref{compp}, \eqref{comppx},
\begin{eqnarray}\label{3ct}
&&N=1,\ a_0=20,\ a_1=a_2=25a_3=-c_1=k_{1}=1, \ x_{01}=0.
\end{eqnarray}
and
\begin{eqnarray}\label{qtanh}
q=10\tanh(y-t),
\end{eqnarray}
at $t=0$. For the partial compacton shown by Fig. 2, the quantity
$G$ possesses the following quite simple form ($u2\equiv u[2]$)
\begin{eqnarray}\label{Pcomp}
u2=\left\{\begin{array}{ll} 0 & |x-t|\geq \frac12\pi,\\
\dfrac{100\cos(x-t){\rm
sech}^2(y-t)}{[(90-10\sin(x-t))+(46-4\sin(x-t))\tanh(y-t)]^2} &
|x-t|<\frac12\pi.
\end{array} \right.
\end{eqnarray}

\input epsf
\begin {figure}
\centering \epsfxsize=7cm\epsfysize=5cm\epsfbox{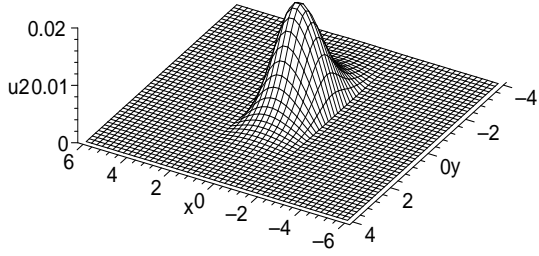}
\caption{Evolution plot for the quantity $u2\equiv u[2]$ expressed
by \eqref{Pcomp} at times $t=0$.}
\end{figure}

Similar to the first kind of compacton (full compacted) solution
as shown in Fig. 1, the detailed study of the interaction between
partial compactons is also non-completely elastic and they will
partially exchange their shapes.

If the arbitrary functions $q_1$ and $p_2$ are not taken as in
\eqref{q1p2} but are solved out from \eqref{w0q1} and
\eqref{v0p2}, then we can obtain some further new types of exact
solutions which can not be obtained by the usual variable
separation approach. Actually, when the arbitrary functions $p_1$
and $q_2$ are fixed by \eqref{v0p1} and \eqref{w0q2} at the same
time, then the corresponding solution(s) for the functions $q_1$
and $p_2$ can be found by solving the \em linear \rm equations
\eqref{v0p2} and \eqref{w0q1}.

In principle, the more kinds of exact solutions can be obtained
from further Darboux transformations starting from the seed
solution \{\eqref{phi2}, \eqref{u2},\ \eqref{v2},\ \eqref{w2}\}.
However, we do not discuss these types of solutions further
because of their complexity.

In summary, the variable separation solutions can be obtained not
only by the truncated Painlev\'e expansion and the generalized
multi-linear equations, but also by other well known approaches
especially by the Darboux transformation. In this short paper, the
Darboux transformation is successfully used to find the variable
separable solutions of the (2+1)-dimensional NNV equation. The
recursive Darboux transformation may yield further new types of
variable separable solutions while the new variable separable
function should satisfy a further constrained condition, say,
\eqref{v0p2} with $v_0$ being given by \eqref{v0p1}.

By selecting the arbitrary functions appropriately, one may obtain
abundant localized excitations like the dromions, lumps, ring
solitons, breathers, instantons, solitoffs, peakons, fractal and
chaotic patterns. In addition to these types of localized
excitations, a further type of the localized excitations,
compactons, is given in this paper. The (2+1)-dimensional
compactons discussed here possess the different type of
interaction properties as that of the ring solitons and peakons.
The interactions among compactons are not completely elastic and
do not exchange their shapes completely.

The authors are in debt to thank the helpful discussions with the
professors X. B. Hu and Y. S. Li and Doctor. C. L. Chen. The work
was supported by the National Outstanding Youth Foundation of
China (No.19925522), the Research Fund for the Doctoral Program of
Higher Education of China (Grant. No. 2000024832) and the Natural
Science Foundation of Zhejiang Province of China.


\begin{thebibliography}{999}
\bibitem{GGKM} C. S. Gardner, J. M. Greene, M. D. Kruskal, R. M. Miura,
Phys. Rev. Lett. {\bf 19}, 1095 (1976).
\bibitem{Hirota} R. Hirota, Phys. Rev. Lett. {\bf 27}, 1192 (1971).
\bibitem{Olver} P. J. Olver,  Application of Lie Group to
Differential Equation, Springer, New York, 1986.
\bibitem{Matveev} V. B. Matveev, M. A. Salle, Darboux Transformations and
Solitons, Springer, Berlin, 1991.
\bibitem{Liu} D. L. Yu, Q. P. Liu, S. K. Wang, J. Phys. A: Math. Gen.  {\bf 35}, 3779 (2002).
\bibitem{Lou} S.-y. Lou, J.-z. Lu, J. Phys. A: Math. Gen.  {\bf
 29}, 4209 (1996).
\bibitem{Miller} W. Miller, Symmetry and Separation of Variables,
                 (Addison-Wesley, Reading, MA, 1977);
E. G. Kalnins, W. Miller, J. Math. Phys.
                            26, 1560 (1985);
               E. G. Kalnins, W. Miller, J. Math. Phys.
                             26, 2168 (1985).
\bibitem{Do1} P. W. Dolye, P. J. Vassiliou, Int. J. Nonlinear
                     Mech. 33, 315 (1998);
              P. W. Dolye, J. Phys. A. 29, 7581 (1996).
\bibitem{Rz1} R. Z. Zhdanov, J. Phys. A 27, L291 (1994);
              R. Z. Zhdanov, I. V. Revenko, W. I. Fushchych, J.
                      Math. Phys. 36, 5506 (1995);
              R. Z. Zhdanov, J. Math. Phys. 38, 1197 (1997).
\bibitem{Zh1} C. Z. Qu, S. L. Zhang and R. C. Liu,  Physica D 144, 97 (2000);
 P. G. Estevez, C. Z. Qu and S. L. Zhang,
              J. Math. Anal. Appl. (2002) in print; S-l Zhang, S-y
              Lou and C-z Qu, Chin. Phys. Lett. (2002) in print.
\bibitem{Zsl} S-l Zhang, S-y Lou and C-z Qu, preprint (2002).
\bibitem{Lou3} C-w Cao, Sci. China A 33, 528 (1990);
Y. Cheng and Y-s Li, Phys. Lett. A 175, 22 (1991);
 B. G. Konopelchenko, V. Sidorenko and W. Strampp, Phys. Lett. A
 175, 17 (1971);
S-y Lou and L-l Chen, J. Math. Phys. 40, 6491 (1999).
\bibitem{Lou1} S-y Lou,  Phys. Lett. A 277(2000) 94.
\bibitem{Lou2} S-y Lou and H.-y. Ruan,  J. Phys. A: Math. Gen. 34, 305 (2001);
              S-y Lou, Physica Scripta, 65, 7 (2002);
             X-y Tang, C-l Chen and S-y Lou, J. Phys. A: Math. Gen. 35, L293 (2002);
             X-y Tang and S-y Lou, Commun. Theor. Phys. 38, 1 (2002);
             S-y Lou, C-l Chen and X-y Tang, J. Math. Phys. 43, 4078 (2002).
\bibitem{Lou4}X-y Tang, S-y Lou and Y. Zhang, Phys. Rev. E. 66, 046601 (2002).
\bibitem{NNV} L. P. Nizhnik, Sov. Phys. Dokl.  {\bf 25}, 706 (1980);
 A. P. Veselov, S. P. Novikov, Sov. Math. Dokl.  {\bf 30}, 588 (1984);
 S. P. Novikov, A. P. Veselov, Physica D  {\bf 18}, 267 (1986).
\bibitem{BLM} M. Boiti, J. J. P. Leon, M. Manna, F. Pempinelli, Inv.
 Problems {\bf 2}, 116 (1986).
\bibitem{Hu1}Y. Tagami, Phys. Lett. A 141, 116 (1989); X-b Hu and
Y-s Li, J. Phys. A: Math. Gen. 24, 1979 (1991).
\bibitem{Hu2}X-b Hu, J. Phys. A: Math. Gen. 27, 1331 (1994).
\bibitem{Radha} R. Radha, M. Lakshmanan, J. Math. Phys. {\bf 35}, 4746 (1994).
\bibitem{LLT} S-y Lou, J. Lin and X-y Tang, Eur. Phys. J. B 22, 473 (2001).
\bibitem{CH}  R. Camassa, and D. D. Holm, Phys. Rev. Lett. 71, 1661 (1993);
F. Calogero, Phys. Lett. A 201, 306 (1995); F. Calogero and J. P.
Francoise, J. Math. Phys. 37, 2863 (1996); Y. B. Suris, Phys.
lett. A217, 321 (1996); B. Fuchssteiner, Physica D95, 229 (1996);
P. J. Olver and P. Rosenau, Phys. Rev. E 53, 1900 (1996); J.
Schiff, Physica D 121, 24 (1998); R. A. Kraenkel, and A. Zenchuk,
Phys. Lett. A 260, 218 (1999); R. A. Kraenkel, M. Senthilvelan and
A. Zenchuk, Phys. Lett. A 273, 183 (2000).
\bibitem{peakon}S-y Lou, preprint (2002).
\bibitem{Rose1}P. Rosenau and J. M. Hyman, Phys. Rev. Lett. 70,
564 (1993).
\bibitem{compacton} F. Cooper, J. M. Hyman, and A. Khare, Phys. Rev. E 64 art. no. 026608
(2001); A. Chertock and D. Levy, J. Comput. Phys. 171, 708 (2001);
M. A. Manna, Physica D 149, 231 (2001); M. Eleftheriou, B. Dey and
G. P. Tsironis, Phys. Rev. E 62, 7540 (2000); P. Tchofo Dinda, T.
C. Kofane and M. Remoissenet, Phys. Rev. E 60, 7525 (1999); P.
Tchofo Dinda and M. Remoissenet, Phys. Rev. E 60, 6218 (1999).
\bibitem{LouWu}S-y Lou and Q-x
Wu, Phys. Lett. A 262, 344 (1999).
\bibitem{Rose2}P. Rosenau,
Phys. Rev. Lett. 73, 1737 (1994).
\end{thebibliography}
\end{document}